\documentstyle[12pt]{article}
\parskip=\medskipamount                 
\thicklines                     
\def\ordless{{\lower2mm\hbox{$\,\stackrel{\textstyle <}{\sim}\, $}}}
\def\ordgt{{\lower2mm\hbox{$\,\stackrel{\textstyle >}{\sim}\, $}}}

\newcommand{\bim}[6]{\bibitem{#1}#2, {\em #3\/}$\;${\bf
#4}$\;$(#5)$\;${#6}.}
\catcode`\@=11

\newif\if@fewtab\@fewtabtrue

\catcode`\@=11

\newif\if@fewtab\@fewtabtrue

{\count255=\time\divide\count255 by 60
\xdef\hourmin{\number\count255}
\multiply\count255 by-60\advance\count255 by\time
\xdef\hourmin{\hourmin:\ifnum\count255<10 0\fi\the\count255}}
\def\ps@draft{\let\@mkboth\@gobbletwo
    \def\@oddhead{}
    \def\@oddfoot
       {\hbox to 7 cm{$\scriptstyle Draft\ version:\ \draftdate$
       \hfil}\hskip -7cm\hfil\rm\thepage \hfil}
    \def\@evenhead{}\let\@evenfoot\@oddfoot}

\def\ceqno{\global\@fewtabfalse
    \ifcase\@eqcnt \def\@tempa{& & &}\or \def\@tempa{& &}
      \or \def\@tempa{&}
      \or\def\@tempa{}\fi\@tempa
{\rm(\theequation)}}

\def\aeqno#1{\global\@fewtabfalse
    \ifcase\@eqcnt \def\@tempa{& & &}\or \def\@tempa{& &}
      \or \def\@tempa{&}
      \or\def\@tempa{}\fi\@tempa
{\rm(\theequation,#1)}}

\def\label#1{\ifnum\draftcontrol=1
 \global\def\draftnote{$\scriptstyle #1$}\fi
 \@bsphack\if@filesw {\let\thepage\relax
   \def\protect{\noexpand\noexpand\noexpand}%
\xdef\@gtempa{\write\@auxout{\string
      \newlabel{#1}{{\@currentlabel}{\thepage}}}}}\@gtempa
   \if@nobreak \ifvmode\nobreak\fi\fi\fi
  \@esphack}

\def\alabel#1#2{\label{#1}\global\@fewtabfalse
    \ifcase\@eqcnt \def\@tempa{& & &}\or \def\@tempa{& &}
      \or \def\@tempa{&}
      \or\def\@tempa{}\fi\@tempa
{\hbox to 3cm{\phantom{\rm(\theequation,#2)}
\draftnote \hfil}\hskip -3cm {\rm(\theequation,#2)}}}

\def\clabel#1{\label{#1}\global\@fewtabfalse
    \ifcase\@eqcnt \def\@tempa{& & &}\or \def\@tempa{& &}
      \or \def\@tempa{&}
      \or\def\@tempa{}\fi\@tempa
{\hbox to 3cm{\phantom{\rm(\theequation)}
\draftnote \hfil}\hskip -3cm{\rm(\theequation)}}}

\def\eqnarray{\def\draftnote{{}}\global\@fewtabtrue
\stepcounter{equation}\let\@currentlabel=\theequation
\global\@eqnswtrue
\global\@eqcnt\z@\tabskip\@centering\let\\=\@eqncr
$$\halign to \displaywidth\bgroup\@eqnsel\hskip\@centering\@eqcnt\z@
  $\displaystyle\tabskip\z@{##}$&\global\@eqcnt\@ne
  \hskip 1\arraycolsep \hfil${##}$\hfil
  &\global\@eqcnt\tw@ \hskip 1\arraycolsep
$\displaystyle\tabskip\z@{##}$
\hfil  \tabskip\@centering&\global\@eqcnt\thr@@\llap{##}\tabskip\z@
\cr}

\def\endeqnarray{\@@eqncr\egroup
      \global\advance\c@equation\m@ne$$\global\@ignoretrue}

\def\@eqnnum{\hbox to 3cm{\phantom{\rm(\theequation)} \draftnote
                         \hfil}\hskip -3cm {\rm(\theequation)}}

\def\@@eqncr{\let\@tempa\relax
    \ifcase\@eqcnt \def\@tempa{& & &}\or \def\@tempa{& &}
      \or \def\@tempa{&}
      \or\def\@tempa{}
\fi\@tempa
\if@eqnsw
\if@fewtab\@eqnnum\fi
\stepcounter{equation}\fi\global
\@eqnswtrue\global\@eqcnt\z@\global\@fewtabtrue\cr}

\def\draftcite#1{\ifnum\draftcontrol=1#1\else{}\fi}

\def\@lbibitem[#1]#2{\item{}\hskip -3cm \hbox to 2cm
{\hfil$\scriptstyle\draftcite{#2}$}\hskip
1cm[\@biblabel{#1}]\if@filesw
     {\def\protect##1{\string ##1\space}\immediate
      \write\@auxout{\string\bibcite{#2}{#1}}}\fi\ignorespaces}

\def\@bibitem#1{\item\hskip -3cm \hbox to 2cm
{\hfil $\scriptstyle\draftcite{#1}$}\hskip 1cm
\if@filesw \immediate\write\@auxout
       {\string\bibcite{#1}{\the\value{\@listctr}}}\fi\ignorespaces}

\def\nsection#1{\section{#1}\setcounter{equation}{0}}

\def\draftdate{\number\month/\number\day/\number\year\ \ \ \hourmin }

\global\def\draftcontrol{0}
\catcode`\@=12

\def\theequation{{\thesection.\arabic{equation}}}

\def\qq{\begin{eqnarray}}
\def\qqq{\end{eqnarray}}

\begin{document}

\begin{titlepage}
\centerline{\hfill                 UTTG--12--93}
\centerline{\hfill                 hep-th/9401060}
\vfill
\begin{center}
{\large \bf Witten's Invariant of 3-Dimensional Manifolds: Loop
Expansion and Surgery Calculus.\footnote{
to be published in the volume {\em Knots and Applications}.}
} \\
\bigskip
\centerline{L. Rozansky\footnote{Work supported
by NSF Grant 9009850 and R. A. Welch Foundation.}}

\centerline{\em Theory Group, Department of Physics, University of
Texas at Austin}
\centerline{\em Austin, TX 78712-1081, U.S.A.}

\vfill
{\bf Abstract}

\end{center}
\begin{quotation}
We review two different methods of calculating Witten's invariant:
a stationary phase approximation
and a surgery calculus. We give a detailed description of the 1-loop
approximation formula for Witten's invariant and of the technics
involved in deriving its exact value through a surgery construction of
a manifold. Finally we compare the formulas produced by both methods
for a 3-dimensional sphere $S^{3}$ and a lens space $L(p,1)$.

\end{quotation}
\vfill
\end{titlepage}

\pagebreak
\tableofcontents
\pagebreak
\nsection{Introduction}

A quantum field theory based on Chern-Simons action has been developed
by E. Witten in his paper \cite{W}. Consider a connection $A_{\mu}$ of
a $G$ bundle $E$ on a 3-dimensional manifold $M$, $G$ being a simple
Lie group. If the bundle is trivial, then an integral
\qq
S_{CS}=\frac{1}{2}\epsilon^{\mu\nu\rho}{\rm Tr}\int_{M}
(A_{\mu}\partial_{\nu}A_{\rho}+
\frac{2}{3}A_{\mu}A_{\nu}A_{\rho})\,d^{3}x.
\label{1.1}
\qqq
defines a Chern-Simons action as a function of $A_{\mu}$. A manifold
invariant $Z(M)$ is a path integral
\qq
Z(M,k)=\int [{\cal D}A_{\mu}]e^{\frac{i}{\hbar}S_{CS}[A_{\mu}]},
\;\;\hbar=\frac{\pi}{k},
\label{1.2}
\qqq
here $k\in{\bf Z}$, and the brackets in $[{\cal D}A_{\mu}]$ mean that
we integrate over the gauge equivalence classes of connections. The
action~(\ref{1.1}) does not depend on the choice of local coordinates
on $M$, neither does it depend on the metric of the manifold $M$.
Therefore the integral $Z(M,k)$, also known to physicists as a
partition function, is a topological invariant of the manifold (modulo
the possible metric dependence of the integration measure $[{\cal
D}A_{\mu}]$).

Witten considered two different methods of calculating $Z(M,k)$. He
first applied a stationary phase approximation to the
integral~(\ref{1.2}). This is a standard method of quantum field
theory. It expresses $Z(M,k)$ as asymptotic series in $k^{-1}$. The
first term in this series contains such ingredients as
Chern-Simons action of flat connections and Reidemeister torsion.
The other method of calculating the invariant, which we call ``surgery
calculus'' is based upon a construction of $M$ as a surgery on a link
in $S^{3}$
(or in $S^{1}\times S^{2}$). It presents $Z(M,k)$ as a finite
sum, however the number of terms in it grows as a power of $k$.
Reshetikhin and Turaev used the surgery calculus formula in~\cite{RT}
as a definition of Witten's invariant and proved its invariance (i.e.
independence of the choice of surgery to construct a given manifold
$M$) without referring to the path integral~(\ref{1.2}).

A systematic comparison between both methods of calculating Witten's
invariant has been initiated in \cite{FG}. D. Freed and R. Gompf
compared the numeric values of the invariants of some lens spaces and
homology spheres for large values of $k$ as given by the two methods.
The full analytic comparison has been carried out in \cite{J} and
\cite{G} for lens spaces and mapping tori. It was extended further to
Seifert manifolds in \cite{R}. A complete agreement between the
stationary phase approximation and surgery calculus has been found in
all these papers.

In this paper we will review both methods of calculating Witten's
invariant and compare their results. In section~\ref{*2} we explain
the stationary phase approximation method. Section~\ref{*3} contains
the basics of surgery calculus. In section~\ref{*4} we apply both
methods to the calculation of Witten's invariant of the sphere
$S^{3}$ and lens space $L(p,1)$.

\nsection{Stationary Phase Approximation}
\label{*2}
\subsection{Finite Dimensional Integrals}
Let us start with a simple example of the stationary phase
approximation. Consider a finite dimensional integral
\qq
Z(\hbar)=\int\frac{d^{n}X}{(2\pi\hbar)^{n/2}}
\exp\left[\frac{i}{\hbar}S(X_{1},\ldots,X_{n})\right]
\label{2.1}
\qqq
for some function $S$. Here $\hbar$ is an arbitrary small constant,
called Planck's constant in quantum theory. The integral~(\ref{2.1})
is a finite dimensional version of the path integral~(\ref{1.2}). Note
that a path integral measure
$[{\cal D}A_{\mu}]$ includes implicitly a
factor $(2\pi\hbar)^{-1/2}=\pi^{-1}(k/2)^{1/2}$ for each of the
one-dimensional integrals comprising the full path integral.

In the limit of small $\hbar$ the dominant contribution to $Z(\hbar)$
comes from the extrema of $S$, i.e. from the points $X^{a}_{i}$ such
that
\qq
\left.\frac{\partial S}{\partial X_{i}}\right|_{X_{i}=X_{i}^{(a)}}=0,
\;\;1\leq i\leq n
\label{2.2}
\qqq

If we retain only the quadratic terms in Taylor expansion of $S$ in
the vicinity of these points, then
\begin{eqnarray}
Z(\hbar)&=&\sum_{a}e^{\frac{i}{\hbar}S(X^{(a)})}
\int d^{n}x\exp\left[i\pi
\sum_{i,j=1}^{n}x_{i}x_{j}
\left.\frac{\partial^{2} S}{\partial X_{i}\partial
X_{j}}\right|_{X_{i}=X_{i}^{(a)}}\right]
\nonumber\\
&=&\sum_{a}e^{\frac{i}{\hbar}S(X^{(a)})}
{\rm det}^{-1/2}
\left(-i\left.\frac{\partial^{2} S}{\partial X_{i}\partial
X_{j}}\right|_{X_{i}=X_{i}^{(a)}}\right).
\label{2.3}
\end{eqnarray}
A phase of this expression requires extra care. The matrix
$\frac{\partial^{2} S}{\partial X_{i}\partial X_{j}}$ is hermitian. It
has only real eigenvalues $\lambda$, but they can be both positive
and negative. Each positive eigenvalue contributes a phase factor
$(-i)^{-1/2}=e^{i\pi/4}$ to the inverse square root of the
determinant in eq.~(\ref{2.3}), while each negative eigenvalue
contributes $i^{-1/2}=e^{-i\pi/4}$. Therefore a refined version of the
formula~(\ref{2.3}) is
\qq
Z(\hbar)=\sum_{a}e^{\frac{i}{\hbar}S(X^{(a)})}
e^{i\frac{\pi}{4}\eta_{a}}
\left|\det\left(\left.\frac{\partial^{2} S}{\partial X_{i}\partial
X_{j}} \right|_{X_{i}=X_{i}^{(a)}}\right)\right|^{-1/2},
\label{2.4}
\qqq
here
\qq
\eta^{a}=\#\,{\rm positive}\,\lambda-\#\,{\rm negative}\,\lambda
\label{2.04}
\qqq

In the context of quantum field theory the integral~(\ref{2.1})
becomes an infinite dimensional path integral, however the stationary
phase approximation method remains the same if we can make sense of
infinite dimensional determinants. Physicists call the
formula~(\ref{2.3}) a 1-loop approximation, because it can be derived
by summing up all 1-loop Feynman diagrams.

\subsection{Gauge Invariant Theories}
The integral~(\ref{1.2}) presents a special challenge, because the
action~(\ref{1.1}) is invariant under a gauge transformation
(i.e. under a local change of basis in the fibers)
\qq
A_{\mu}\rightarrow A_{\mu}^{g}=g^{-1}A_{\mu}g+g^{-1}\partial_{\mu}g.
\label{2.5}
\qqq
The integral over the gauge equivalence classes of connections is
equal to the integral over all connections divided by the volume of
the group of gauge transformations. However the latter integral can
not be calculated through eq.~(\ref{2.4}), because its stationary
phase points are not isolated. They form the orbits of the gauge
action~(\ref{2.5}). Therefore we should rather integrate over the
submanifold in the space of all connections, which is transversal to
gauge orbits, multiply the terms in the sum~(\ref{2.4}) by the volumes
of those orbits and divide the whole sum by the volume of the group of
gauge transformations.

The problem of reducing an integral of a function invariant under the
action of a group, to an integral over a factor manifold, is not
unfamiliar to mathematicians. For example, an integral of the product
of characters over a simple Lie group can be reduced to its maximal
torus at a price of adding an extra factor which accounts for the
volume of the orbits of adjoint action. This factor is equal to the
square of denominator in the Weyl character formula and appears as a
Jacobian of a certain coordinate transformation.

A similar trick was developed for gauge invariant path integrals by
Faddeev and Popov. Consider a Lie algebra valued functional
$\Phi[A_{\mu}]$ such that each gauge orbit intersects transversally
the set of its zeros
\qq
\Phi[A_{\mu}]=0
\label{2.7}
\qqq
For a constant function $g(x)=g={\rm const}$, the second term in
eq.~(\ref{2.5}) vanishes. If $g$ also belongs to the center $Z(G)$ of
$G$, then, obviously, $A^{g}=A$. Therefore a general gauge orbit
intersects the set~(\ref{2.7}) at the same point ${\rm Vol}\,(Z(G))$
times. ${\rm Vol}\,(Z(G))$ denotes the number of elements in $Z(G)$.
We use this notation to make connection with the formula~(\ref{2.30}),
which, as we will see, works also for the case when the tangent spaces
of the manifold~(\ref{2.7}) and a gauge orbit intersect along a finite
dimensional space.

A path integral generalization of a simple formula
\qq
\int_{-\infty}^{+\infty}\delta(f(x))\frac{dx}{\sqrt{2\pi\hbar}}=
\sum_{x_{i}:f(x_{i})=0}\left|\sqrt{2\pi\hbar}
f^{\prime}(x_{i})\right|^{-1}
\label{2.8}
\qqq
can be used to derive the following identity:
\qq
1=\frac{1}{{\rm Vol}\,(Z(G))}
\left|\det\left(\left.\frac{\delta(\sqrt{2\pi\hbar}\Phi[A^{g}])}
{\delta g}\right|_{\Phi[A^{g}]=0}\right)\right|
\int{\cal D}g\,\delta(\Phi[A^{g}]).
\label{2.9}
\qqq
A path integral $\delta$-function which will reduce the integral over
all connections to a submanifold~(\ref{2.7}) is called ``gauge
fixing''.

A multiplication of the integral~(\ref{1.2}) by the r.h.s. of the
identity~(\ref{2.9}) and a subsequent change of variables
$A_{\mu}^{g}\rightarrow A_{\mu}$ allows us to factor the volume of the
group of gauge transformations out of the integral over all gauge
connections:
\begin{eqnarray}
Z(\hbar)&=&\left({\cal D}g\right)^{-1}\int{\cal
D}A_{\mu}\,e^{\frac{i}{\hbar}S_{CS}[A_{\mu}]}
\nonumber\\
&=&\left({\cal D}g\right)^{-1}
\frac{1}{{\rm Vol}\,(Z(G))}
\int{\cal D}A_{\mu}\,e^{\frac{i}{\hbar}S_{CS}[A_{\mu}]}
\left|\det\left(\left.\frac{\delta(\sqrt{2\pi\hbar}\Phi[A^{g}])}
{\delta g}\right|_{\Phi[A^{g}]=0}\right)\right|
\int{\cal D}g\,\delta(\Phi[A^{g}])
\nonumber\\
&=&
\frac{1}{{\rm Vol}\,(Z(G))}
\int{\cal D}A_{\mu}\,e^{\frac{i}{\hbar}S_{CS}[A_{\mu}]}
\delta(\Phi[A])
\left|\det\left(\left.\frac{\delta(\sqrt{2\pi\hbar}\Phi[A^{g}])}
{\delta g}\right|_{g=1}\right)\right|.
\label{2.11}
\end{eqnarray}
%

\subsection{Chern-Simons Path Integral}
Let us apply a stationary phase approximation to the
integral~(\ref{2.11}). We first look for the stationary phase
points\footnote{Eq.(\ref{2.12}) can be
used to verify the gauge invariance of the
action~(\ref{1.1}) under small gauge transformations. The
infinitesimal version of eq.~(\ref{2.5}) is
%
\[
\delta_{\omega}A_{\mu}=D_{\mu}\omega,
\]
%
so that
\begin{eqnarray}
\delta_{\omega}S&=&\int\frac{\partial S}{\partial A_{\mu}}
D_{\mu}\omega\,d^{3}x\sim\epsilon^{\mu\nu\rho}\,{\rm Tr}\int
F_{\nu\rho}D_{\mu}\omega\,d^{3}x
\nonumber\\
&=&-{\rm
Tr}\int\omega\epsilon^{\mu\nu\rho}D_{\mu}F_{\nu\rho}\,d^{3}x=0
\nonumber
\end{eqnarray}
because of Bianchi identity.}
\qq
\frac{\delta S}{\delta A_{\mu}}\sim\epsilon^{\mu\nu\rho}F_{\nu\rho}=0
\label{2.12}
\qqq
These points are flat connections, i.e. connections for which
$F_{\mu\nu}=0$. The gauge equivalence classes of flat connections are
in one-to-one correspondence with the homomorphisms
\qq
\pi_{1}({\cal M})\stackrel{A}{\rightarrow}G,\;\;A:\;x\mapsto g(x)\in G
\label{2.13}
\qqq
up to a conjugacy, that is, the homomorphisms
$x\rightarrow g(x)$ and $x\rightarrow h^{-1}g(x)h$ are considered
equivalent.

The next step is to expand the action~(\ref{1.1}) up to the terms
quadratic in gauge field variation $a_{\mu}$ around a particular flat
connection $A_{\mu}^{(a)}$:
\qq
S_{CS}[A_{\mu}^{(a)}+\pi\sqrt{\frac{2}{k}}a_{\mu}]\approx
S_{CS}[A_{\mu}^{(a)}]+\frac{\pi^{2}}{k}\epsilon^{\mu\nu\rho}
{\rm Tr}\int a_{\mu}D_{\nu}a_{\rho}\,d^{3}x.
\label{2.14}
\qqq
Here $D_{\nu}$ is a covariant derivative with respect to the
``background field'' $A_{\mu}^{(a)}$:
\qq
D_{\nu}=\partial_{\nu}+[A_{\nu}^{(a)},\ast].
\label{2.014}
\qqq

A gauge fixing condition should be imposed on the fluctuation field
$a_{\mu}$. Witten suggested a covariant (with respect to
$A_{\mu}^{(a)}$)
choice\footnote{
Note that $\Phi[a_{\mu}]$ depends on the metric of $M$. This
dependence ultimately results in a framing dependence of $Z(M,k)$.
}:
\qq
\Phi[a_{\mu}]=D_{\mu}a_{\mu}.
\label{2.15}
\qqq
According to eq.~(\ref{2.5}), a change of $a_{\mu}$ under an
infinitesimal gauge transformation  $g(x)\approx 1+\omega(x)$ is
\qq
\delta_{\omega}a_{\mu}=\frac{1}{\pi}\sqrt{\frac{k}{2}}D_{\mu}\omega.
\label{2.16}
\qqq
Therefore the operator
$\left.\delta(\sqrt{2\pi\hbar}\Phi[A^{g}])/\delta g\right|_{g=1}$ is a
covariant Laplacian $\Delta=D_{\mu}D_{\mu}$ acting on 0-forms. As for
the $\delta$-function of $\Phi[a_{\mu}]$, it can be presented as a
path integral over a Lie algebra valued scalar field:
\qq
\delta(\Phi[a_{\mu}])=\int{\cal D}\phi
\exp\left[2\pi i\,
{\rm Tr}\int\phi D_{\mu}a_{\mu}\,d^{3}x\right].
\label{2.17}
\qqq
As a result, at the 1-loop level
\begin{eqnarray}
Z(M,k)&\approx&\frac{1}{{\rm Vol}\,(Z(G))}\sum_{a}
e^{i\frac{k}{\pi}S_{CS}[A_{\mu}^{(a)}]}|\det\Delta|
\nonumber\\
&&\times\int{\cal D}a_{\mu}{\cal D}\phi\exp\left[
i\pi\,{\rm Tr}\int d^{3}x\,
(\epsilon^{\mu\nu\rho}a_{\mu}D_{\nu}a_{\rho}+
2\phi D_{\mu}a_{\mu})\right]
\nonumber\\
&=&\frac{1}{{\rm
Vol}\,(Z(G))}\sum_{a} e^{i\frac{k}{\pi}S_{CS}[A_{\mu}^{(a)}]}
\frac{|\det\Delta|}{(\det L_{-})^{1/2}}.
\label{2.18}
\end{eqnarray}
Here $L_{-}$ is the operator of the quadratic form in the exponential
of the path integral. $L_{-}$ acts on the direct sum of 0-forms and
1-forms on $M$:
\qq
L_{-}(\phi,a_{\mu})=(D_{\mu}a_{\mu},
\epsilon_{\mu\nu\rho}D_{\nu}a_{\rho}-D_{\mu}\phi).
\label{2.19}
\qqq
If we use 3-forms instead of 0-forms, then $L_{-}=\star D+D\star$,
$\star$ is the Hodge operator. Note that the integration measures
of the fluctuation fields ${\cal D}\phi$ and ${\cal D}a_{\mu}$ do not
contain any implicit factors in contrast to ${\cal D}g$ and ${\cal
D}A_{\mu}$.

\subsection{$\eta$-Invariant}
A. Schwartz observed in \cite{S} that the absolute value of the ratio
of determinants in eq.~(\ref{2.18}) was equal to the square root of
the Reidemeister-Ray-Singer analytic torsion. The phase of the ratio
is equal to the $\eta$-invariant of Atiyah, Patodi and Singer.
Similarly to eq.~(\ref{2.04}) it is a difference between the number of
positive and negative eigenvalues of $L_{-}$. Thus the formula for the
ratio of determinants is
\qq
\frac{|\det\Delta|}{(\det
L_{-})^{1/2}}=\tau_{R}^{1/2}e^{i\frac{\pi}{4}\eta}.
\label{2.20}
\qqq
Actually $L_{-}$ has infinitely many eigenvalues, so a
regularization is needed to define $\eta$. To get some idea of how
$\eta$ might depend on the background connection $A_{\mu}^{(a)}$
consider the following simple problem. Let the eigenvalues be
$\lambda_{m}=m+a,\;m\in{\bf Z}$ and let us calculate the number of
positive $\lambda_{m}$ minus the number of negative ones as a function
of $a$. The simplest regularization is
\qq
\eta_{a}=\lim_{\epsilon\rightarrow 0}
\left[\sum_{\lambda_{m}>0}e^{-\lambda_{m}\epsilon}-
\sum_{\lambda_{m}<0}e^{\lambda_{m}\epsilon}\right].
\label{2.21}
\qqq
Suppose that $0<a<1$, then
\qq
\eta_{a}=\lim_{\epsilon\rightarrow 0}\left[
\frac{e^{-a\epsilon}}{1-e^{-\epsilon}}-
\frac{e^{-(1-a)\epsilon}}{1-e^{-\epsilon}}\right]=1-2a.
\label{2.22}
\qqq
In particular $\eta_{1/2}=0$ because of the symmetry
between the positive and negative $\lambda$ for $a=1/2$.
A dependence of $\eta_{a}$ on $a$ is a nontrivial consequence of the
infinity of the number of eigenvalues, since naively the number of
positive and negative $\lambda$ is the same for any $a\in(0,1)$.

Obviously, $\eta_{a}$ is a periodic function of $a$:
$\eta_{a+n}=\eta_{a},\;n\in{\bf Z}$, because $a$ and $a+n$ define the
same set of eigenvalues $\lambda$. Therefore eq.~(\ref{2.22}) requires
a modification to work for all $a$. Indeed, when $a$ moves through an
integer number $n$, an eigenvalue $\lambda_{-n}$ changes the sign and
the value of $\eta_{a}$ jumps by two units. Define $I_{a}$ to be a
number of positive eigenvalues becoming negative minus a number of
negative eigenvalues becoming positive when the parameter goes from
$1/2$ to $a$. Then
\qq
\eta_{a}=1-2a-2I_{a}.
\label{2.33}
\qqq
If we define $\eta_{0}$ to be equal to 1 as if $\lambda_{0}=0$ is
counted as positive, then we arrive to the formula
\qq
\eta_{a}=\eta_{0}-1+(1-2a)-2I_{a}.
\label{2.24}
\qqq
A similar formula for the $\eta$-invariant of $L_{-}$ was derived in
\cite{FG}:
\qq
\eta_{a}=\eta_{0}-(1+b^{1}(M)){\rm
dim}G+\frac{4}{\pi^{2}}c_{V}S_{CS}[A_{\mu}^{(a)}]-2I_{a},
\label{2.25}
\qqq
here $\eta_{0}$ is the $\eta$-invariant of the trivial connection,
$b^{1}(M)$ is the first Betti number of $M$, $I_{a}$ is a spectral
flow of $L_{-}$ and $c_{V}$ is a dual Coxeter number of the group $G$
(e.g. $c_{V}=N$ for $SU(N)$).
The operator $L_{-}$ for a trivial connection has
${\rm dim}G$ 0-form zero modes which are constant Lie algebra valued
functions and $b^{1}(M){\rm dim}G$ 1-form zero modes which are Lie
algebra valued closed 1-forms. All these modes are counted as positive
in $\eta_{0}$, hence the term $(1+b^{1}(M)){\rm dim}G$. The role of
the smooth function $1-2a$ is played by
$\frac{4}{\pi^{2}}c_{V}S_{CS}[A_{\mu}^{(a)}]$.

The metric of $M$ enters the gauge fixing functional~(\ref{2.15}) as
well as the operators $\Delta$ and $L_{-}$. We could naively assume
that this dependence would cancel out from the ratio of determinants
in eq.~(\ref{2.20}). However the phase $\eta$ has an ``anomalous''
dependence on the metric. It can be compensated by multiplying
$Z(M,k)$ by an extra phase factor
\qq
\exp\left[-i{\rm dim}G\frac{1}{96\pi}\int_{M}{\rm Tr}
(\omega\wedge d\omega+
\frac{2}{3}\omega\wedge\omega\wedge\omega)d^{3}x\right] ,
\label{2.26}
\qqq
here $\omega$ is a Levi-Chivita connection on $M$ and the integral in
the exponent is its Chern-Simons invariant. This invariant is defined
relative to the choice of basis in the tangent space at each point of
$M$. The local change in that basis is the analog of the gauge
transformation. The exponent of eq.~(\ref{2.26}) is invariant under
the transformations which are homotopic to identity. The choice of
basis modulo such transformations is called ``framing''. The change in
framing by $n$ units shifts the phase of the factor~(\ref{2.26}) by
$\pi n\,{\rm dim}G/12$. Actually the whole invariant~(\ref{2.1}) with
a compensated metric dependence would be multiplied by a factor
\qq
\exp\left[i\frac{\pi}{12}n\,{\rm dim}G\frac{k}{k+c_{V}}\right].
\label{2.026}
\qqq

Physicists call the exponent of eq.~(\ref{2.26}) a 1-loop
counterterm.
It converts the metric dependence of $\eta$ into a framing dependence
of the invariant $Z(M,k)$. According to~\cite{FG},
\qq
\eta_{0}=0
\label{2.0126}
\qqq
in the special framing of $M$ called canonical.

\subsection{Zero Modes}
To complete the study of the stationary phase approximation we have to
consider the flat connections for which the operators $L_{-}$ and
$\Delta$ have zero modes. The 0-form zero modes of these operators
satisfy the same equation
\qq
D_{\mu}\omega=0,
\label{2.0226}
\qqq
so they are the elements
of a cohomology $H_{a}^{0}$ built upon a covariant
derivative~(\ref{2.014}). A 1-form $a_{\mu}$ which is a zero mode of
$L_{-}$, satisfies two equations

\qq
\epsilon^{\mu\nu\rho}D_{\nu}a_{\rho}=0,\;\;D_{\mu}a_{\mu}=0
\label{2.27}
\qqq
The first equation means that $a_{\mu}$ is a closed form with respect
to $D$, the second one means that it is not exact: if
$a_{\mu}=D_{\mu}\omega$, then $D^{2}_{\mu}\omega=0$, hence
$D_{\mu}\omega=0$. As a result, the 1-form zero modes are the elements
of the cohomology $H_{a}^{1}$.

Let us remove the zero modes from the operators $L_{-}$ and $\Delta$.
The absolute value of the ratio of their determinants is still equal
to the square root of the Reidemeister torsion, which, as noted
in~\cite{J}, becomes an element of
$\Lambda^{\rm max}H_{a}^{0}\otimes(\Lambda^{\rm
max}H_{a}^{1})^{\ast}$. As for the phase $\eta$, it can be obtained by
a simple correction of eq.~(\ref{2.25}) presented in~\cite{FG}:
\qq
\eta_{a}=\eta_{0}-(1+b^{1}(M)){\rm dim}G-
({\rm dim}H_{a}^{0}+{\rm dim}H_{a}^{1})+
\frac{4}{\pi^{2}}c_{V}S_{CS}[A_{\mu}^{(a)}]-2I_{a}.
\label{2.28}
\qqq
The zero modes of $L_{-}$ are counted as positive in the spectral flow
$I_{a}$. Therefore their number had to be subtracted from $\eta_{a}$
since they are removed from the l.h.s. of eq.~(\ref{2.20}) and do not
affect its phase.

According to eq.~(\ref{2.0226}), the 0-form zero modes are the
infinitesimal gauge transformations that do not change the background
field $A_{\mu}^{(a)}$. The group of gauge transformations which is a
symmetry of $A_{\mu}^{(a)}$, is isomorphic to a subgroup $H_{a}\subset
G$ which commutes with the image of the homomorphism~(\ref{2.13}).
Therefore $H_{a}^{0}$ is isomorphic to a Lie algebra of $H_{a}$.

The 1-form zero modes $a_{\mu}$ are the deformations of a flat
connection $A_{\mu}^{(a)}$ which preserve its flatness in the linear
order in $a_{\mu}$:
\qq
F_{\mu\nu}[A_{\rho}^{(a)}+a_{\rho}]\approx
\epsilon_{\mu\nu\lambda}\epsilon^{\lambda\sigma\rho}
D_{\sigma}a_{\rho}=0.
\label{2.29}
\qqq
In most cases these infinitesimal deformations can be extended up to
the finite flatness preserving deformations. Then $H_{a}^{1}$ is a
tangent space of the moduli space ${\cal M}_{a}$ of flat connections
at the point $A_{\mu}^{(a)}$.

A removal of the 0-form zero modes from the determinants of the r.h.s.
of eq.~(\ref{2.20}) amounts to ``forgetting'' about $H_{a}$ as a part
of the group of gauge transformations. In other words, the
symmetry under the global $H_{a}$ gauge transformations\footnote{
a gauge transformation~(\ref{2.5}) is called global if the
transformation parameter
$g(x)$ is (covariantly) constant.} is not
fixed, as it is demonstrated on a simple finite dimensional example in
the Appendix of~\cite{R}. As a result, we have to divide the integrals
of eq.~(\ref{2.11}) by the volume of $H_{a}$ ``by hands''. A square
root of the Reidemeister torsion as an element of
$\Lambda^{\rm max}H_{a}^{0}\otimes
\left(\Lambda^{\rm max}H_{a}^{1}\right)^{\ast}$
defines a
``ratio'' of the volume forms on ${\cal M}_{a}$ and $H_{a}$.
Therefore $\sqrt{\tau_{R}}/{\rm Vol}(H_{a})$ is a volume form on
${\cal M}_{a}$ and it is quite natural to supplement a sum in
eq.~(\ref{2.18}) by an integral over the components of the moduli
space.

The volume forms for $H_{a}$ and ${\cal M}_{a}$ being the part of the
path integral measure, contain the factors
$(2\pi\hbar)^{-1/2}=\pi(k/2)^{1/2}$. After extracting these factors
from the integration measures we obtain the following 1-loop formula:
\begin{eqnarray}
Z(M,k)&=&\sum_{a}e^{\frac{i}{\pi}(k+c_{V})S_{CS}[A_{\mu}^{(a)}]}
e^{-i\frac{\pi}{4}[(1+b^{1}(M)){\rm dim}G+
{\rm dim}H^{0}_{a}+{\rm dim}H_{a}^{1}+2I_{a}]}
\left(\frac{k}{2\pi^{2}}\right)^
{\frac{{\rm dim}H^{1}_{a}-{\rm dim}H_{a}^{0}}{2}}
\nonumber\\
&&\times\frac{1}{{\rm Vol}\,(H_{a})}\int_{{\cal M}_{a}}\tau_{R}^{1/2}.
\label{2.30}
\end{eqnarray}
The sum goes over the connected components of the moduli space of flat
connections on $M$. The Chern-Simons action
$S_{CS}[A_{\mu}^{(a)}]$ is constant within those components, because
its derivative is zero due to eq.~(\ref{2.12}).

The formula~(\ref{2.30}) is not the end of the story, because
sometimes ${\rm dim}\,{\cal M}_{a}<{\rm dim}\,H_{a}^{1}$. In other
words, not all the 1-form zero modes of $L_{-}$ can be extended to
finite deformations of the flat connection $A_{\mu}^{(a)}$. This means
that the Chern-Simons action is not constant in the direction of these
modes, rather its expansion around $A_{\mu}^{(a)}$ starts with the
terms of order $m>2$. The corresponding piece of the path integral has
a form
\qq
\int d^{n}x\,\exp\left[2\pi i(2\pi\hbar)^{\frac{m-2}{2}}
\frac{1}{m!}\left.\frac{\partial^{(m)}S}
{\partial X_{i_{1}}\ldots
\partial X_{i_{m}}}\right|_{X_{i}=X_{i}^{(a)}}
x_{i_{1}}\cdot\ldots\cdot x_{i_{m}}\right]
\sim e^{i\pi\frac{n}{m}}(2\pi\hbar)^{\frac{n(2-m)}{2m}},
\label{2.31}
\qqq
here $n={\rm dim}\,H_{a}^{1}-{\rm dim}\,{\cal M}_{a}$ and
$\hbar=\pi/k$ as defined in eq.~(\ref{1.2}). Therefore if
${\rm dim}\,{\cal M}_{a}<{\rm dim}\,H_{a}^{1}$, then we should
substitute ${\rm dim}\,{\cal M}_{a}$ instead of ${\rm dim}\,H_{a}^{1}$
in the r.h.s. of eq.~(\ref{2.30}) and multiply it by the
factor~(\ref{2.31}).

\nsection{Surgery Calculus}
\label{*3}
\subsection{Multiplicativity in Quantum Theory}
The surgery calculus uses one of the basic principles of quantum field
theory: a multiplicativity of the path integral~(\ref{1.2}). We are
going to describe briefly what this multiplicativity means. Suppose
that a 3-dimensional manifold $M$ has a boundary $\partial M$. Let us
impose a boundary condition on a connection $A_{\mu}$. For example, we
choose a tangent vector field $v_{\mu}$ on $\partial M$ and demand
that $A_{\mu}v_{\mu}$ is equal to some fixed function $A$ on $\partial
M$:
\qq
A_{\mu}v_{\mu}=A.
\label{3.1}
\qqq
Then a path integral~(\ref{1.2}) taken over all the connections on $M$
satisfying this condition becomes a functional $\Psi[A]$. Such
functional is called a wave function or a state in quantum theory. All
possible functionals $\Psi[A]$ for a given manifold $\partial M$ form
a Hilbert space ${\cal H}_{\partial M}$. A scalar product in it is
defined by the path integral over all functions $A$ on $\partial M$:
\qq
\langle\Psi_{2}|\Psi_{1}\rangle=
\int \bar{\Psi_{2}}[A]\Psi_{1}[A]{\cal D}A.
\label{3.2}
\qqq

Suppose now that two manifolds $M_{1}$ and $M_{2}$ have diffeomorphic
boundaries (with opposite orientations): $\partial M_{1}=\partial
M_{2}$. We can glue them together to form a single manifold $M$. An
integration over connections $A_{\mu}$ on $M$ can be split into an
integration over connections $A_{\mu}^{(1)}$ on $M_{1}$ and
connections $A_{\mu}^{(2)}$ on $M_{2}$ satisfying the same
condition~(\ref{3.1}) and an integration over all boundary conditions
$A$. If
$\left.A_{\mu}^{(1)}v_{\mu}\right|_{\partial M_{1}}=
\left.A_{\mu}^{(2)}v_{\mu}\right|_{\partial M_{2}}$,
then the Chern-Simons
action is additive\footnote{
In fact, the action~(\ref{1.1}) on a manifold with a boundary should
be corrected by a certain boundary term which guarantees that a
derivative transversal to $\partial M$ does not act on a tangential
component of $A_{\mu}$ which is not fixed by condition~(\ref{3.1}) and
hence is not necessarily continuous after the gluing.}:
\qq
S_{M}[A_{\mu}]=S_{M_{1}}[A_{\mu}^{(1)}]
+S_{M_{2}}[A_{\mu}^{(2)}].
\label{3.3}
\qqq
Since the exponential is multiplicative, the integrals over
$A_{\mu}^{(1)}$ and $A_{\mu}^{(2)}$ can be calculated
separately yielding the wave functions $\Psi_{1,2}[A]$. The whole
integral is a product $\Psi_{1}[a]\bar{\Psi_{2}}[A]$ ($\Psi_{2}$ is
complex conjugated because $\partial M_{1}$ and $\partial M_{2}$ have
opposite orientations). The final integral over $A$ gives a scalar
product:
\qq
Z(M,k)=\langle\Psi_{2}|\Psi_{1}\rangle.
\label{3.4}
\qqq
To summarize, multiplicativity means that gluing the manifolds is
achieved by taking a scalar product of the states appearing on their
boundaries.

We adopt the strategy of \cite{RT}. Each 3-dimensional manifold can be
constructed by a surgery on a link in $S^{3}$. The tubular
neighborhoods of the link components are cut out, the modular
transformations on their boundaries are performed
and then they are glued back. So if we find the wave functions on both
sides of the boundaries of tubular neighborhoods, then we can use
eq.~(\ref{3.4}) to find Witten's invariant.
The boundaries of the tubular neighborhoods are 2-dimensional tori
$T^{2}$, so we start by describing the Hilbert space
${\cal H}_{T^{2}}$. We use canonical quantization as described
in~\cite{EMSS}, where it was called ``first constraining, then
quantizing''.

\subsection{Canonical Quantization}
Consider a manifold
$M={\bf\rm R}^{1}\times T^{2}={\bf\rm R}^{1}
\times S^{1}\times S^{1}$ with
coordinates $t$ along ${\bf\rm R}^{1}$ and $x_{1,2}$ along both
circles, $0\leq x_{1,2}<1$. The Chern-Simons action~(\ref{1.1}) can be
cast in the form (up to a total derivative in $t$ that can be removed
by adding appropriate boundary terms):
\qq
S_{CS}={\rm Tr}\int
dt\,d^{2}x\,(A_{2}\partial_{0}A_{1}+ A_{0}F_{12}).
\label{3.5}
\qqq
The 1-form $A_{\mu}$ takes values in the Lie algebra of $G$. Lie
algebra elements are antihermitian matrices in the adjoint
representation. Quantum field theory deals usually with hermitian
objects, so we introduce hermitian forms
\qq
\tilde{A}_{\mu}=-iA_{\mu},\;\;\tilde{F}_{\mu\nu}=-iF_{\mu\nu}.
\label{3.05}
\qqq
Now
\qq
S_{CS}=-{\rm Tr}\int dt\,d^{2}x\,
(\tilde{A}_{2}\partial_{0}\tilde{A}_{1}+
\tilde{A}_{0}\tilde{F}_{12}).
\label{3.0015}
\qqq
Compare this with the action of a constrained mechanical system
\qq
S=\int dt\,\left[p_{i}\dot{q}_{i}+h(p_{i},q_{i})+
\lambda_{\alpha}\phi_{\alpha}(p_{i},q_{i})\right],
\label{3.6}
\qqq
here $q_{i}$ are coordinates, $p_{i}$ are conjugate momenta,
$h(p_{i},q_{i})$ is a hamiltonian, $\phi_{\alpha}(p_{i},q_{i})$ are
constraints and $\lambda_{\alpha}$ are Lagrange multipliers. We see
that $\tilde{A}_{1}$ and $-\tilde{A}_{2}$ are conjugate coordinates
and momenta. The hamiltonian is zero as it happens in diffeomorphism
invariant theories.

A path integral over $A_{0}$ in eq.~(\ref{1.2}) produces a
$\delta$-function of the constraint $F_{12}$, so we should, in fact,
study only flat 2-dimensional connections as coordinates in the
phase space. A gauge transformation can make both $A_{1}$ and $A_{2}$
constant. Moreover, since $\pi_{1}(T^{2})$ is commutative, $A_{1}$ and
$A_{2}$ will belong to the same Cartan subalgebra
(e.g. they will be
made diagonal simultaneously for $G=SU(N)$). The action~(\ref{3.0015})
becomes simply
\qq
S_{CS}=\int dt \tilde{A}_{2}^{a}\dot{\tilde{A}}_{1}^{a},
\label{3.7}
\qqq
an index $a$ runs over the orthonormal basis of Cartan subalgebra.
After a quantization the fields $\tilde{A}_{i}^{a}$ become
hermitian operators $\hat{\tilde{A}}_{i}^{a}$ satisfying the Heisenberg
commutation relation:
\qq
[\hat{\tilde{A}}_{2}^{a},\hat{\tilde{A}}_{1}^{b}]=
i\hbar\delta^{ab}\equiv i\frac{\pi}{k}\delta^{ab}.
\label{3.8}
\qqq
This algebra can be represented in a space of functions
$\psi(\tilde{A}_{1}^{a})$:
\qq
\hat{\tilde{A}}_{1}^{a}\psi(\tilde{A}_{1}^{b})=
\tilde{A}_{1}^{a}\psi(\tilde{A}_{1}^{b}),\;\;
\hat{\tilde{A}}_{2}^{a}\psi(\tilde{A}_{1}^{b})=
i\hbar\frac{\partial}{\partial\tilde{A}_{1}^{a}}
\psi(\tilde{A}_{1}^{b}).
\label{3.9}
\qqq
The eigenfunctions of $\hat{\tilde{A}}_{1}^{a}$ are
$\delta$-functions, while the eigenfunctions of
$\hat{\tilde{A}}_{2}^{a}$ are exponentials
\qq
|\alpha_{a};2\rangle\sim e^{i\alpha_{a}\tilde{A}_{1}^{a}},
\label{3.09}
\qqq
here we use a standard quantum mechanical notation for eigenstates:
\qq
\hat{\tilde{A}}_{i}^{a}|\alpha_{b};i\rangle=
\hbar\alpha^{a}|\alpha_{b};i\rangle.
\label{3.10}
\qqq
However a construction of a representation for the algebra
$\hat{\tilde{A}}_{i}^{a}$ should reflect the fact that a Cartan
subalgebra is not an appropriate configuration space for the torus
$T^{2}$.

\subsection{$U(1)$ Theory}
Let us study carefully the simplest case of $G=U(1)$. The constant
field components appearing in the action~(\ref{3.7}) are equal to the
contour integrals along the periods $C_{1,2}$ of $T^{2}$
\qq
\tilde{A}_{i}=\oint_{C_{i}}\tilde{A}_{j}(x)\,dx^{j}
\label{3.010}
\qqq
of any connection $\tilde{A}_{j}(x)$ which can be reduced to a
constant one by a homotopically trivial gauge transformation. A
homotopically nontrivial gauge transformation
\qq
g(x)=e^{2\pi i(m_{1}x_{1}+m_{2}x_{2})}
\label{3.11}
\qqq
is well defined if $m_{1,2}\in {\bf Z}$. Eq.~(\ref{2.5}) shows that
$\tilde{A}_{1}$ and $\tilde{A}_{2}$ remain constant under this
transformation, but their values are shifted:
\qq
\tilde{A}_{i}\rightarrow \tilde{A}_{i}+2\pi m_{i}.
\label{3.12}
\qqq
Thus both coordinate $\tilde{A}_{1}$ and momentum $\tilde{A}_{2}$ are
periodic with a period of $2\pi$. The phase space is compact (it is
$S^{1}\times S^{1}$), its volume is $(2\pi)^{2}$ and, according to the
WKB approximation, the dimension of the Hilbert space should be
approximately
\qq
{\rm dim}{\cal H}_{T^{2}}^{U(1)}\approx
\frac{(2\pi)^{2}}{2\pi\hbar}\equiv 2k
\label{3.13}
\qqq
in the limit of large $k$. In fact, as we will see, eq.~(\ref{3.13})
is exact.

A periodicity in $\tilde{A}_{1}$ leads to a quantization of the
eigenvalues of $\hat{\tilde{A}}_{2}$: $\alpha$ should be integer to
make the eigenfunctions~(\ref{3.09}) periodic. On the other hand,
since $\tilde{A}_{2}$ is also periodic, we should limit the number of
independent values of $\alpha$, e.g.
\qq
-k\leq\alpha<k,\;\;\alpha\in{\bf Z}
\label{3.14}
\qqq
This procedure is self-consistent, because for integer $k$ the period
of $\tilde{A}_{2}$ ($2\pi$) is a multiple of the spacing of its
eigenvalues ($\hbar=\pi/k$).

The $2k$ values of $\alpha$ determine the momentum eigenstates
$|\alpha;2\rangle$, which form an orthonormal basis of
${\cal H}_{T^{2}}^{U(1)}$. Another basis is formed by the coordinate
eigenstates $|\alpha;1\rangle$ with the same range~(\ref{3.14}) of
possible values of $\alpha$. These two bases are related by a finite
dimensional version of the Fourier transform (which also provides a
relation between coordinate and momentum eigenstates in quantum
mechanics of a particle on a line):
\qq
|\alpha;2\rangle=\frac{1}{\sqrt{2k}}
\sum_{\beta=-k}^{k-1}e^{-i\frac{\pi}{k}\alpha\beta}
|\beta;1\rangle.
\label{3.15}
\qqq
%

\subsection{Modular Transformations}
A unimodular transformation of cycles $C_{1,2}$ in eq.~(\ref{3.010})
generates a canonical transformation of our system:
\qq
C_{i}\stackrel{U}{\mapsto}C_{i}^{\prime}=U_{ij}C_{j},\;
\tilde{A}_{i}\stackrel{U}{\mapsto}\tilde{A}_{i}^{\prime}=
U_{ij}\tilde{A}_{j},\; U\in SL(2,{\bf Z}).
\label{3.014}
\qqq
Therefore $SL(2,{\bf Z})$ can be represented in
${\cal H}_{T^{2}}^{U(1)}$. This group is generated by two elements
\qq
S=\left(
\begin{array}{cc}
0&-1\\
1&0\end{array}\right),\;\;
T=\left(
\begin{array}{cc}
1&1\\0&1\end{array}\right)
\label{3.015}
\qqq
satisfying a relation
\qq
(ST)^{3}=S^{2}
\label{3.16}
\qqq
Each matrix
\qq
U^{(p,q)}=\left(\begin{array}{cc}p&r\\q&s\end{array}\right)
\in SL(2,{\bf Z})
\label{3.17}
\qqq
can be presented as their product
\qq
U^{(p,q)}=T^{a_{t}}S\ldots T^{a_{1}}S.
\label{3.18}
\qqq
The integer numbers $a_{i}$ form a continued fraction expansion of
$p/q$:
\qq
\frac{p}{q}=a_{t}-\frac{1}{a_{t-1}-\frac{1}{\ldots-\frac{1}{a_{1}}}}.
\label{3.19}
\qqq
We denote as $\hat{U}^{(p,q)}$ an action of $U^{(p,q)}$ in
${\cal H}_{T^{2}}^{U(1)}$. According to eq.~(\ref{3.18}), it is
determined by choosing $\hat{S}$ and $\hat{T}$.

The matrix $S$
interchanges coordinate and momentum operators:
\qq
\hat{S}\hat{\tilde{A}}_{1}\hat{S}^{-1}=\hat{\tilde{A}}_{2},\;\;
\hat{S}\hat{\tilde{A}}_{2}\hat{S}^{-1}=-\hat{\tilde{A}}_{1}.
\label{3.20}
\qqq
The same is achieved by the matrix of eq.~(\ref{3.15}), so
\qq
\hat{S}_{\alpha\beta}=\frac{1}{\sqrt{2k}}e^{-i\frac{\pi}{k}\alpha\beta}
\label{3.21}
\qqq
in the coordinate basis $|\alpha;1\rangle$. We use a formula
\qq
\exp\left[\frac{ik}{2\pi}\hat{\tilde{A_{1}^2}}\right]
\hat{\tilde{A}}_{2}
\exp\left[-\frac{ik}{2\pi}\hat{\tilde{A_{1}^2}}\right]=
\hat{\tilde{A}}_{2}+\hat{\tilde{A}}_{1},
\label{3.22}
\qqq
which is easy to check by using a representation~(\ref{3.9}), in order
to find $\hat{T}$ in the coordinate basis:
\qq
\hat{T}_{\alpha\beta}=e^{-i\frac{\pi}{12}}
e^{\frac{i\pi}{2k}\alpha^{2}}\delta_{\alpha\beta}
\label{3.23}
\qqq
The phase of $\hat{T}$ is chosen to comply with eq.~(\ref{3.16}).

\subsection{$SU(2)$ Theory}
Let us turn now to the case of $G=SU(2)$. Its Cartan subalgebra (an
algebra of diagonal traceless antihermitian $2\times 2$ matrices) is
isomorphic to that of $U(1)$. A new feature is the Weyl reflection. A
global gauge transformation
\qq
g=\left(\begin{array}{cc}
0&1\\
-1&0\end{array}\right)
\label{3.24}
\qqq
changes the signs of $\tilde{A}_{1}$ and $\tilde{A}_{2}$:
\qq
g\tilde{A}_{i}g^{-1}=-\tilde{A}_{i},\;\;i=1,2.
\label{3.25}
\qqq
The phase space should be factored by this transformation, its volume
becoming half of that for $U(1)$. Therefore we expect a dimension of
the Hilbert space ${\cal H}_{T^{2}}^{SU(2)}$ also to be approximately
half of that of ${\cal H}_{T^{2}}^{U(1)}$.

The space ${\cal H}_{T^{2}}^{SU(2)}$ is isomorphic to a subspace of
${\cal H}_{T^{2}}^{U(1)}$ which is antisymmetric under the Weyl
reflection~(\ref{3.25}), if we make an identification
\qq
k_{U(1)}=k_{SU(2)}+2=K_{SU(2)}.
\label{3.26}
\qqq
In other words, the basis of ${\cal H}_{T^{2}}^{SU(2)}$ is formed by
the antisymmetric combinations
\qq
|\alpha;i\rangle_{SU(2)}=\frac{1}{\sqrt{2}}
\left(|\alpha;i\rangle_{U(1)}-|-\alpha;i\rangle_{U(1)}\right),\;
0<\alpha<K,\;\;\;
{\rm dim}\,{\cal H}_{T^{2}}^{SU(2)}=K-1.
\label{3.27}
\qqq
As a result, the matrices
$\hat{S}_{\alpha\beta}$ and $\hat{T}_{\alpha\beta}$ for $SU(2)$
are obtained (after a minor change in phase factors) by restricting
the matrices~(\ref{3.21}) and (\ref{3.23}) to the
subspace~(\ref{3.27}):
\qq
\hat{S}_{\alpha\beta}=\sqrt{\frac{2}{K}}\sin\frac{\pi\alpha\beta}{K},\;\;
\hat{T}_{\alpha\beta}=e^{-\frac{i\pi}{4}}
e^{\frac{i\pi}{2K}\alpha^{2}}\delta_{\alpha\beta}
\label{3.28}
\qqq
Eq.~(\ref{3.18}) was used in~\cite{J} in order to derive a formula for
$\hat{U}^{(p,q)}$:
\qq
\hat{U}^{(p,q)}_{\alpha\beta}= -i\frac{{\rm sign}(q)}{\sqrt{2K|q|}}
e^{-\frac{i\pi}{4}\Phi(M^{(p,q)})}\sum_{\mu=\pm
1}\sum_{n=0}^{q-1}\mu\exp\frac{i\pi}{2Kq}\left[p\alpha^{2}+
2\mu\alpha(\beta+2Kn)+s(\beta+2Kn)^{2}\right]
\label{3.29}
\qqq
Here $\Phi(M)$ is a Rademacher phi function defined as follows
\qq
\Phi\left[\begin{array}{cc}p&r\\q&s\end{array}\right]=\left\{
\begin{array}{lll}
\frac{p+s}{q}-12s(s,q)&\;{\rm if}&\;q\neq0\\
\frac{r}{s}&\;{\rm if}&\;q=0\end{array}\right.,
\label{3.30}
\qqq
a function $s(s,q)$ being a Dedekind sum:
\qq
s(m,n)=\frac{1}{4n}\sum_{j=1}^{n-1}\cot\frac{\pi j}{n}\cot\frac{\pi
mj}{n}.
\label{3.31}
\qqq
%

\subsection{A General Simple Lie Group}
Consider now a general simple Lie group $G$. A gauge transformation can
make the fields $A_{i}$ constant and belonging to a Cartan subalgebra
of the Lie algebra associated with $G$. The homotopically nontrivial
gauge transformations like~(\ref{3.11}) force the eigenvalues $\alpha$
of the eigenvectors $|\alpha;i\rangle$ of $\hat{\tilde{A}}_{i}$ to
belong to the weight lattice $\Lambda_{w}$ of $G$ factored by the root
lattice $\Lambda_{R}$ magnified $K$ times (here $K=k+c_{V}$). The Weyl
reflections similar to~(\ref{3.25}) require us to take only the Weyl
antisymmetric combinations
\qq
\sum_{w\in W}(-1)^{|w|}|w(\alpha);i\rangle,
\label{3.32}
\qqq
here $W$ is the Weyl group and $|w|$ denotes a determinant of the
transformation $w$.
As a result, the basis of ${\cal H}_{T^{2}}^{G}$ is formed by the
states~(\ref{3.32}) ($i$ is either 1 or 2), for which
$\alpha\in\Lambda_{w}$ belongs to the fundamental domain of the affine
Weyl group $\tilde{W}$. This group is a semidirect product of the Weyl
group $W$ and a group of translations by the elements of the lattice
$K\Lambda_{R}$. The walls of the fundamental domain should be
excluded, that is, we require $\tilde{w}(\alpha)\neq\alpha$ for any
$\tilde{w}\in\tilde{W}$.

A scalar product $\langle\alpha;i|\beta;i\rangle$
of the basis elements of ${\cal H}_{T^{2}}^{G}$  is equal to 1 if
$\alpha$ and $\beta$ are the shifted highest weights of conjugate
representations, otherwise it is zero.

The formulas for $\hat{S}$ and $\hat{T}$ matrices of the simply
laced Lie group G, as presented in \cite{J} (see also~\cite{K}), are
\begin{eqnarray}
\hat{S}_{\alpha\beta}=i^{|\Delta_{+}|}
\left|\frac{{\rm Vol}\,\Lambda_{w}}
{{\rm Vol}\,K\Lambda_{R}}\right|^{1/2}
\sum_{w\in W}(-1)^{|w|}\exp\left(-\frac{2\pi i}{K}
\langle w(\alpha),\beta\rangle\right),\\
\hat{T}_{\alpha\beta}=\delta_{\alpha\beta}
\exp\left(\frac{i\pi}{K}\langle\alpha,\alpha\rangle-
\frac{i\pi}{c_{V}}\langle\rho,\rho\rangle\right),
\label{3.34}
\end{eqnarray}
here $\Delta_{+}$ is a set of positive roots of $G$, $|\Delta_{+}|$ is
their number, $\langle\;\;,\;\;\rangle$ is a Cartan scalar product
normalized so that the length of all roots of $G$ is equal to
$\sqrt{2}$, $\rho=\frac{1}{2}\sum_{\alpha\in\Delta_{+}}\alpha$. Note
that $\hat{S}_{\alpha\beta}$ is proportional to the numerator of the
Weyl character formula.

Now we should explain how to produce the states $|\alpha;i\rangle$ by
taking a path integral over a 3-dimensional manifold $M$ with a
boundary $\partial M=T^{2}$. First we extend the path
integral~(\ref{1.2}) by adding the so-called Wilson lines. Consider a
closed manifold $M$ with a link $L$ inside it. Let us attach
representations $V_{\alpha_{i}}$ of $G$ to its components $L_{i}$.
Here $V_{\alpha}$ denotes a representation of $G$ with the shifted
highest weight $\alpha$ (i.e. the highest weight of $V_{\alpha}$
is $\alpha-\rho$). For any connection $A_{\mu}$ on $M$ the trace of
its holonomy
\qq
{\cal O}_{i}={\rm Tr}_{V_{\alpha_{i}}}{\rm P\,exp}\,\oint A_{\mu}\,
dx^{\mu}
\label{3.35}
\qqq
is invariant under gauge and coordinate transformations. Therefore the
path integral
\qq
Z^{\alpha_{1},\ldots,\alpha_{n}}(M,L,k)=
\int[{\cal D}A_{\mu}]e^{i\frac{k}{\pi}S_{CS}(A_{\mu})}
\prod_{i}{\cal O}_{i}
\label{3.36}
\qqq
is an invariant of the link $L$ in $M$. Witten used the methods of
conformal field theory to prove that it satisfies skein relations.
Thus he proved that this invariant is equal to the Jones
polynomial up to a normalization constant.

Take the integral~(\ref{3.36}) for a solid
torus $M=S^{1}\times D^{2}\;\:(\partial M=S^{1}\times\partial
D^{2}=T^{2})$ and $L$ consisting of one component $S^{1}\times P$, $P$
being the center of the disk $D^{2}$. Let $C_{1}$ be the cycle which
is contractible through $M$. Witten claimed in \cite{W}, that if we
attach a representation $V_{\alpha}$ to the only component of $L$,
then the integral~(\ref{3.36}) produces a state $|\alpha;1\rangle$ in
${\cal H}_{T^{2}}^{G}$. In particular, a solid torus without any
link inside it is equivalent to a torus with a link carrying a trivial
representation, hence it has a state $|\rho;1\rangle$ on its boundary.

Let us cut out a tubular neighborhood of an $n$-component link $L$ in
$S^{3}$. A remaining piece $S^{3}\setminus L$ has a boundary
$(T^{2})^{n}$. Therefore a path integral~(\ref{1.2}) taken over
$S^{3}\setminus L$ produces a state $| L\rangle$ in
$({\cal H}_{T^{2}}^{G})^{\otimes n}$. Suppose that we glue a tubular
neighborhood back
after putting a Wilson line~(\ref{3.35}) inside each of its
components. Then according to the multiplicativity law~(\ref{3.4}),
\qq
Z^{\alpha_{1},\ldots,\alpha_{n}}(M,L,k)=
\langle L|\,\bigotimes_{i=1}^{n}|\alpha_{i};1\rangle
\label{3.036}
\qqq
Therefore the state $|L\rangle$ can be identified with the tensor
$Z^{\alpha_{1},\ldots,\alpha_{n}}$ belonging to the dual Hilbert space
$\left[({\cal H}_{T^{2}}^{G})^{\otimes n}\right]^{\ast}$.
The latter can be
calculated by using cabling (for the link components which carry
representations of $G$ other than the fundamental one) and skein
relations.

Let us glue the components of the tubular neighborhood of $L$ back
after performing modular transformations $U^{(p_{i},q_{i})}$ on their
boundaries. Any 3-dimensional manifold $M$ can be constructed in this
way. A multiplicativity law~(\ref{3.4}) leads to the following
expression for its invariant:
\qq
Z(M,K)=\sum_{\alpha_{1},\ldots,\alpha_{n}}
Z^{\alpha_{1},\ldots,\alpha_{n}}(S^{3},L,k)
\hat{U}_{\alpha_{1}\rho}^{(p_{1},q_{1})}\ldots
\hat{U}_{\alpha_{n}\rho}^{(p_{n},q_{n})}.
\label{3.37}
\qqq
The sum here goes, of course, over $\alpha_{i}$  belonging to the
fundamental domain of the affine Weyl group $\tilde{W}$. Different
surgeries on different knots in $S^{3}$ can produce the same manifold
$M$. Reshetikhin and Turaev proved in \cite{RT} that the
value of the r.h.s. of
eq.~(\ref{3.37}) is the same for all such surgeries.

\nsection{Some Examples}
\label{*4}
\subsection{A Gluing Formula}
We are going to use the surgery calculus in order to calculate the
$U(1)$ and $SU(2)$ invariants of some simple 3-dimensional manifolds
$M$. We will construct these manifolds by gluing together 2 solid tori
after performing a modular transformation $U\in SL(2,{\bf
Z})$ on the surface of one of them. Since neither of the tori has a
Wilson line~(\ref{3.35}) inside it, then they have a state
$|\rho;1_{i}\rangle\in{\cal H}_{T^{2}}^{G}$ corresponding to
a trivial representation, on their boundary.  An index $i=1,2$
refers to
the fact that there are two contractible cycles $C_{1}^{(i)}$ on the
common boundary $T^{2}$: $C_{1}^{(1)}$ is contractible through the
first solid torus while $C_{1}^{(2)}$ is contractible through the
second one. Let us use the basis in the Hilbert space ${\cal
H}_{T^{2}}^{G}$ corresponding to the cycles $C_{i}^{(1)}$ of the
first solid torus, then
\qq
|\rho,1_{2}\rangle=\sum_{\alpha}\hat{U}_{\rho\alpha}
|\alpha;1_{1}\rangle.
\label{4.01}
\qqq
According to the multiplicativity law~(\ref{3.4}), Witten's invariant
of the manifold constructed by gluing the tori, is a scalar product
\qq
\langle\rho;1_{1}|\rho;1_{2}\rangle=\hat{U}_{\rho\rho}.
\label{4.02}
\qqq
We will calculate the matrix element $U_{\rho\rho}$ with the help of
eqs.~(\ref{3.21}), (\ref{3.23}) and~(\ref{3.28}).  The
gluing induces a particular framing of the manifold, which may differ
from the canonical one, so we will supplement $\hat{U}_{\rho\rho}$
with a correction factor~(\ref{2.026}). Then we will compare its large
$k$ limit with the stationary phase approximation
formula~(\ref{2.30}). Since $U(1)$ is abelian, its Chern-Simons action
is purely quadratic. Therefore the path integral~(\ref{1.2}) is
gaussian and the formula~(\ref{2.30}) should be exact for the $U(1)$
invariant.

\subsection{3-Dimensional Sphere}
We start with the 3-dimensional sphere $S^{3}$. The two solid tori
that form it are glued through a modular transformation $S$. The
induced framing is canonical, so the invariants are
\begin{eqnarray}
Z_{U(1)}(S^{3},k)&=&\hat{S}_{00}=\frac{1}{\sqrt{2k}},
\label{4.1}\\
Z_{SU(2)}(S^{3},k)&=&\hat{S}_{11}=
\sqrt{\frac{2}{K}}\sin\frac{\pi}{K}
\stackrel{k\rightarrow\infty}{\longrightarrow}
\sqrt{2}\pi K^{-3/2}.
\label{4.2}
\end{eqnarray}
To determine the invariants entering eq.~(\ref{2.30}) we note that
$\pi_{1}(S^{3})$ is trivial and so is the only flat connection on
$S^{3}$. Its Chern-Simons action is zero and $\tau_{R}=1$. All the
phase factors of eq.~(\ref{2.30}) can be dropped due to
eq.~(\ref{2.0126}). For any $U(1)$ flat connection on a manifold with
$b_{1}(M)=0$
\qq
H_{a}=U(1),\;{\rm dim}\,H_{a}^{0}={\rm dim}\,H_{a}=1,\;
{\rm dim}\,H_{a}^{1}=0,
\label{4.3}
\qqq
while for the trivial $SU(2)$ connection
\qq
H_{a}=SU(2),\;{\rm dim}\,H_{a}^{0}={\rm dim}\,H_{a}=3,\;
{\rm dim}\,H_{a}^{1}=0
\label{4.4}
\qqq
As a result eq.~(\ref{4.1}) and the r.h.s. of eq.~(\ref{4.2}) coincide
with the 1-loop formula~(\ref{2.30}) if we assume that
\qq
{\rm Vol}\,(U(1))=2\pi,\;\;{\rm Vol}\,(SU(2))=2\pi^{2}.
\label{4.5}
\qqq
Both volumes are perfectly consistent since $SU(2)$ is a 3-dimensional
sphere and $U(1)$ is its big circle.

\subsection{A Lens Space $L(p,1)$}
A less trivial example of a manifold is a lens space $L(p,1)$. It is
constructed by gluing two solid tori through a modular transformation
\qq
U^{(-p,1)}=ST^{-p}S.
\label{4.6}
\qqq
According to~\cite{FG} and \cite{J}, the induced framing differs from
the canonical one by $p-3$ units, so in canonical framing
\qq
Z_{U(1)}(L(p,1),k)=e^{-i\frac{\pi}{12}(p-3)}
(\hat{S}\hat{T}^{-p}\hat{S})_{00}=
e^{i\frac{\pi}{4}}\frac{1}{2k}\sum_{\alpha=0}^{2k-1}
\exp\left(-\frac{i\pi}{2k}p\alpha^{2}\right).
\label{4.7}
\qqq
We changed here the range of summation over $\alpha$
from~(\ref{3.14}) to an equivalent one $0\leq\alpha<2k$.

The fundamental group of $L(p,1)$ is ${\bf Z}_{p}$, so there are $p$
flat $U(1)$ connections corresponding to different
homomorphisms~(\ref{2.13}). Therefore our objective is to transform
the r.h.s. of eq.~(\ref{4.6}) into a sum of $p$ terms. We are going to
use a Poisson resummation formula, which relates a sum over integer
numbers of a function and its Fourier transform:
\qq
\sum_{\alpha\in{\bf Z}}f(\alpha)=\sum_{m\in{\bf Z}}
\int_{-\infty}^{+\infty}e^{2\pi im\alpha}f(\alpha)\,d\alpha
\label{4.8}
\qqq
The sum in eq.~(\ref{4.7}) has a finite range, but we can extend it by
using a periodicity of its summand as a function of integer numbers:
\qq
\exp\left[-\frac{i\pi}{2k}p(\alpha+2kn)^{2}\right]=
\exp\left[-\frac{i\pi}{2k}p\alpha^{2}\right],\;\;
{\rm for}\;\alpha,n\in{\bf Z}.
\label{4.9}
\qqq
A ``regularization'' formula
\qq
\sum_{\alpha=0}^{T-1}f(\alpha)=\lim_{\epsilon\rightarrow 0}
(T\epsilon^{1/2})\sum_{\alpha\in{\bf Z}}
e^{-\pi\epsilon\alpha^{2}}f(\alpha),\;\;
{\rm if}\;f(\alpha+T)=f(\alpha)\;{\rm for}\;\alpha,T\in{\bf Z}
\label{4.10}
\qqq
together with eq.~(\ref{4.8}) allow us to reexpress the
invariant~(\ref{4.6})
\qq
Z_{U(1)}(L(p,1),k)=
\frac{e^{i\frac{\pi}{4}}}{2k}\lim_{\epsilon\rightarrow 0}
(2k\epsilon^{1/2})\sum_{m\in{\bf Z}}\int_{-\infty}^{+\infty}
e^{-\pi\epsilon\alpha^{2}}\exp i\pi
\left[-\frac{1}{2k}p\alpha^{2}+2m\alpha\right].
\label{4.11}
\qqq
The integral over $\alpha$ is gaussian, it is exactly equal to
the contribution of the stationary phase point
\qq
\alpha_{m}=2\frac{k}{p}m
\label{4.12}
\qqq
determined by eq.~(\ref{2.3}). The only effect of the prefactor
$e^{-\pi\epsilon\alpha^{2}}$ to the leading order in $\epsilon$ is to
suppress that contribution by the factor
$e^{-\pi\epsilon\alpha_{m}^{2}}$:
\qq
Z_{U(1)}(L(p,1),k)=\lim_{\epsilon\rightarrow 0}(2k\epsilon^{1/2})
\sum_{m\in{\bf Z}}e^{-\pi\epsilon\alpha_{m}^{2}}
\frac{1}{\sqrt{2kp}}\exp\left(\frac{i\pi}{2k}p\alpha_{m}^{2}\right).
\label{4.13}
\qqq
The stationary phase points $\alpha_{m}$ and their contributions
exhibit the same symmetry under the action of the affine Weyl group,
as the original summand in eq.~(\ref{4.7}). This means that if we
add $p$ to $m$, then $\alpha_{m}$ is shifted by $2k$, while the last
exponential of eq.~(\ref{4.13}) remains unchanged. Therefore we can
roll eq.~(\ref{4.10}) backwards in order to limit the summation range
of $m$ to its fundamental domain $0\leq m<p$:
\qq
Z_{U(1)}(L(p,1),k)=\sum_{m=0}^{p-1}\frac{1}{\sqrt{2kp}}
\exp\left(2\pi ik\frac{m^{2}}{p}\right).
\label{4.14}
\qqq
Thus we conclude that the 1-loop contributions of stationary
points~(\ref{4.12}) appear to be exact. We just have to limit the sum
to those $\alpha_{m}$ which belong to the fundamental domain of
$\alpha$:
\qq
0\leq\alpha_{m}<2k.
\label{4.014}
\qqq

We achieved our goal of resumming eq.~(\ref{4.7}). Now a
comparison with the stationary phase formula~(\ref{2.30}) is
straightforward. The Chern-Simons action of the flat $U(1)$ connection
is known to be
\qq
S_{m}=2\pi^{2}\frac{m^{2}}{p},
\label{4.0014}
\qqq
its Reidemeister torsion is $1/p$. We can again drop all the phase
factors of eq.~(\ref{2.30}) since in this case all $\eta_{m}=0$.
Eqs.(\ref{4.3}) and (\ref{4.5}) complete the picture: the 1-loop
formula~(\ref{2.30}) is really exact.

A calculation of the $SU(2)$ invariant of the lens
space $L(p,1)$ goes along
the similar lines. The invariant in the canonical framing is equal to
\begin{eqnarray}
&Z_{SU(2)}(L(p,1),k)=\exp\left[-i\frac{\pi}{4}(p-3)\frac{K-2}{K}\right]
\sum_{\alpha=1}^{K-1}
\hat{S}_{1\alpha}\hat{T}^{-p}_{\alpha\alpha}\hat{S}_{\alpha 1}&
\nonumber\\
&=-\frac{1}{2K}\exp\frac{i\pi}{4K}\left[2p+3(K-2)\right]
\sum_{\alpha=1}^{K-1}\sum_{\mu_{1},\mu_{2}=\pm 1}\mu_{1}\mu_{2}
\exp-\frac{i\pi}{2K}\left[p\alpha^{2}+2\alpha(\mu_{1}+\mu_{2})\right].
&
\label{4.15}
\end{eqnarray}
Again we apply a Poisson resummation formula~(\ref{4.8}). We limit the
sum over the stationary phase points~(\ref{4.12}) to those which
belong to the $SU(2)$ affine Weyl group fundamental domain:
\qq
0\leq\alpha_{m}\leq K.
\label{4.16}
\qqq
Recall that
it is twice as small as that of $U(1)$, because the $SU(2)$ affine
Weyl group includes a reflection $\alpha\rightarrow -\alpha$. A
contribution of the stationary phase points which lie on the
boundaries of the domain~(\ref{4.16}) should be cut in half (in the
case of $U(1)$ we could avoid this by excluding the point $\alpha=2K$
from the domain~(\ref{4.014})). If $p$ is odd, then there is only one
such point $\alpha_{0}=0$, and the resummed expression~(\ref{4.15}) is
\begin{eqnarray}
Z_{SU(2)}(L(p,1),k)&=&-i\sqrt{\frac{2}{Kp}}\exp\frac{i\pi}{2K}(p-3)
\nonumber\\
&&\times
\left[\frac{1}{2}\left(e^{\frac{2\pi i}{Kp}}-1\right)+
\sum_{m=1}^{\frac{p-1}{2}}
\left(e^{\frac{2\pi i}{Kp}}\cos\frac{4\pi m}{p}-1\right)\,
\exp\left(2\pi iK\frac{m^{2}}{p}\right)\right]
\nonumber\\
&\stackrel{k\rightarrow\infty}{\longrightarrow}&
\sqrt{2}\pi(Kp)^{-3/2}
+\sum_{m=1}^{\frac{p-1}{2}}\frac{i}{\sqrt{2Kp}}
\left(2\sin\frac{2\pi m}{p}\right)^{2}\,\exp\left(2\pi
iK\frac{m^{2}}{p}\right).
\label{4.17}
\end{eqnarray}
The number of terms in this equation is approximately half of that in
the $U(1)$ formula~(\ref{4.14}). The number of flat $SU(2)$
connections is also approximately twice as small as that of $U(1)$,
because the Weyl reflection (a conjugation by  $g\in SU(2)$ of
eq.~(\ref{3.24})) makes the nontrivial homomorphisms
$\pi_{1}(L(p,1))={\bf Z}_{p}\rightarrow U(1)$ pairwise equivalent.

The first term of the r.h.s. of eq.~(\ref{4.17}) is a contribution of
the trivial connection. Indeed, the Reidemeister torsion of the
trivial connection is $p^{-3}$, so we get an agreement with
eq.~(\ref{2.30}). The sum in the r.h.s. of eq.~(\ref{4.17}) goes over
nontrivial flat connections. Their Chern-Simons action is again given
by eq.~(\ref{4.0014}). Since $\pi_{1}(L(p,1))={\bf Z}_{p}$ is
abelian, it is mapped by the homomorphism~(\ref{2.13}) into
$U(1)\subset SU(2)$, so that its image commutes with the group
$H_{m}=U(1)$. Therefore
\qq
{\rm dim}\,H_{m}^{0}={\rm dim}\,H_{m}=1.
\qqq
It is also known that ${\rm dim}\,H_{m}^{1}=0$. According to \cite{FG}
and \cite{J},
\qq
\exp\left(-i\frac{\pi}{2}I_{m}\right)=-i.
\qqq
Combining all the pieces we see that the formula~(\ref{2.30})
coincides with the r.h.s. of eq.~(\ref{4.17}). The 1-loop formula is
again demonstrated to work properly.

\nsection{Discussion}
\label{*5}
Despite an obvious progress in calculating and understanding Witten's
invariant, many open questions still remain. The $1/k$ expansion of
knot invariants carried through the Feynman diagram technics in
\cite{ALR}, \cite{GMM} and \cite{DBN} was very successful.
The terms in this expansion appeared to be Vassiliev knot invariants,
and they are expressed as integrals generalizing in a certain way the
gaussian linking number. However, a systematic loop expansion of the
manifold invariants started in~\cite{AS}, proved to be technically
hard. At the same time, the exact resummed formulas for lens spaces
(like the middle expression in eq.~(\ref{4.17})), which are the sums
over flat connections, look abnormally simple and nice from the
quantum field theory point of view. Moreover, the contributions of the
irreducible flat connections on Seifert manifolds, extracted from the
surgery formulas in \cite{R}, are finite loop exact (that is, the
corrections to the terms of eq.~(\ref{2.30}) go only up to a finite
order in $1/k$ expansion). All these facts require a genuine
3-dimensional explanation.

Another possible development of the quantum Chern-Simons theory was
suggested by Witten in~\cite{W1}. He noted that if the fermionic gauge
fields were properly added to the action~(\ref{1.1}) (in other words,
if the Chern-Simons theory was based on an appropriate supergroup),
then their determinant might cancel the bosonic one (i.e. a square
root of the Reidemeister torsion) in eq.~(\ref{2.30}) up to a sign. A
resulting invariant would be a sum over flat connections, each taken
with a certain sign. Witten conjectured that Casson invariant might
be obtained in that way. A Chern-Simons invariant based on a
supergroup $U(1|1)$ was studied in \cite{RS}. It is related to the
Alexander polynomial and also produces a ``$U(1)$ Casson invariant''
which is simply the order of the homology group. It is possible that a
generalization of this theory to other supergroups, such as $U(2|2)$
may produce Casson invariant and provide a quantum field theory
explanation for its calculation through surgery construction (see e.g.
\cite{WA}).
\section*{Acknowledgements}
I am thankful to C. DeWitt-Morette, D. Freed, L. Kauffman
and A. Vaintrob for inviting me to give review talks at their seminars
and for encouraging me to write these notes. I am also indebted to
H. Saleur for many useful discussions.

\end{document}